\documentstyle{article}
\date{\today}

\begin{document}
\title{The $\Sigma_c$ and $\Lambda_c$ 
magnetic moment from QCD spectral Sum Rules}
\author{{Shi-lin Zhu,$^1$ W-Y. P. Hwang,$^{2,3}$ and Ze-sen Yang$^1$}\\
{$^1$Department of Physics, Peking University, Beijing, 100871, China}\\
{$^2$Department of Physics, National Taiwan University, Taipei, 
Taiwan 10764}\\
{$^3$Center for Theoretical Physics, Laboratory for Nuclear Science 
       and}\\
{Department of Physics, Massachusetts Institute of Technology, Cambridge,}\\ 
{Massachusetts 02139}
}
\maketitle

\begin{center}
\begin{minipage}{120mm}
\vskip 0.6in
\begin{center}{\bf Abstract}\end{center}
{The QCD spectral sum rules 
in the presence of the external electromagnetic 
field $F_{\mu\nu}$ is used to calculate the 
magnetic moment of $\Sigma_c$ and $\Lambda_c$. Our result is  
$\mu_{\Sigma^{++}_c}=(2.1\pm 0.3) \mu_N$,
$\mu_{\Sigma^{+}_c}=(0.6\pm 0.1) \mu_N$, 
$\mu_{\Sigma^{0}_c}=(-1.6\pm 0.2) \mu_N$ 
and $\mu_{\Lambda_c}=(0.15\pm 0.05)\mu_N$.\\ 
{\large Keywords: heavy baryon, magnetic moment, QCD sum rule }\\
{\large PACS: 13.40.Em, 14.20.Lq, 14.20.Mr, 12.38.Lg}\\
}
\end{minipage}
\end{center}

\large

\vskip 1.5cm
\par
QCD spectral sum rules (QSSR) \cite{SVZ} are successful in extracting the masses and 
coupling constants of low-lying mesons and baryons. In this approach the 
nonperturbative effects are taken into account through various condensates 
in the QCD vacuum. As shown in \cite{yung,io} the light baryon masses are 
determined by the chiral symmetry breaking quark condensate. In the infinite
heavy quark mass limit the QSSR was first used to evaluate the heavy baryon 
mass \cite{shuryak}. The QSSR with finite heavy quark mass are treated in 
\cite{vmb,bag1}. Recently the QSSR is employed in the framework of the 
heavy quark effective theory \cite{agg,bag2,dai,pc,korner}.

\par
The baryon magnetic moment is another important static quantity as the baryon 
mass. Ioffe and Smilga \cite{IOFFE}, independently, Balitsky and Yung 
\cite{Balit}, extracted the nucleon magnetic moment treating the 
electromagnetic field $F_{\mu\nu}$ as an external field in the QSSR approach. 
They found that the nucleon magnetic moment is essentially related to the 
quark condensate and three susceptibilities $\chi$, $\kappa$ and $\xi$. 
Later on the octet baryon magnetic moments \cite{Chiu} were obtained  
in a similar manner. In this work we shall employ the 
same approach to calculate the 
magnetic moments of the $\Sigma_c$ and $\Lambda_c$.

\par
We shall consider the two-point correlator 
$\Pi_{\Sigma_c} (p)$ in the presence of an external 
electromagnetic field $F_{\alpha\beta}$. 
\begin{equation}
\begin{array}{ll}
\Pi_{\Sigma_c} (p)=& i\int d^4 x \langle 0|T \{ \eta_{\Sigma_c}(x), 
{\bar \eta}_{\Sigma_c} (0) \} 
|0\rangle_{F_{\alpha\beta}} e^{ip\cdot x}  \\
&=\Pi_0(p) + \Pi_{\mu\nu} (p) F^{\mu\nu} +\cdots 
\end{array}
\end{equation} 
where $\Pi_0(p)$ is the polarization operator without the external field 
$F_{\alpha\beta}$. 
The $\eta_{\Sigma_c}$ in Eq. (1)
is the interpolating current with $\Sigma_c$ quantum numbers. 
\begin{equation}
\eta_{\Sigma_c}(x) =\epsilon^{abc}  \{
[{u^a}^T (x) Cc^b(x) ] u^c(x) 
-
[{u^a}^T (x) C\gamma_5 c^b(x) ]\gamma_5 u^c(x) \}
\, ,
\end{equation}
and 
\begin{equation}
\langle 0| \eta_{\Sigma_c}(0) |\Sigma_c \rangle =\lambda_{\Sigma_c} 
\nu_{\Sigma_c} (p) \, ,
\end{equation}
where $u^a(x)$ and $c^b(x)$ is the up and charm quark field, 
$\lambda_{\Sigma_c}$ is the overlap amplititude of the 
interpolating current with the baryon state, 
and the $\nu_{\Sigma_c}$ is the Dirac spinor of the heavy quark. 

\par
Three different tensor structures contribute to $\Pi_{\mu\nu}(p)$,
\begin{equation}
\Pi_{\mu\nu}(p) =\Pi (p) (\sigma_{\mu\nu} {\hat p} +{\hat p}\sigma_{\mu\nu})
+\Pi_1(p) i(p_{\mu}\gamma_{\nu}-p_{\nu}\gamma_{\mu}){\hat p} 
+\Pi_2(p) \sigma_{\mu\nu}  \, .
\end{equation}
As in the original QSSR analysis of nucleon magnetic moment \cite{IOFFE}, 
we shall consider the first tensor structure 
$(\sigma_{\mu\nu} {\hat p} +{\hat p}\sigma_{\mu\nu})$. 

\par
The presence of the external field $F_{\mu\nu}$ will induce 
three new condensates in the QCD vacuum \cite{IOFFE}.
\begin{equation} 
\begin{array}{c}
\langle 0 | {\overline  q} \sigma_{\mu\nu} q |0 \rangle_{F_{\mu\nu}} = e_c  \chi
F_{\mu\nu} \langle 0 | {\overline  q}  q |0 \rangle \, , \\ 
g_s \langle 0 | {\overline  q} {{\lambda^n}\over {2}}G^n_{\mu\nu} q |0 \rangle_{F_{\mu\nu}}
= e_c  \kappa F_{\mu\nu} 
\langle 0 | {\overline  q}  q |0 \rangle \, , \\
g_s \epsilon^{\mu\nu\lambda\sigma} 
\langle 0 | {\overline  q} \gamma_5 {{\lambda^n}\over {2}} G^n_{\lambda\sigma} q |0 \rangle_{F_{\mu\nu}}
= i e_c  \xi F^{\mu\nu} 
\langle 0 | {\overline  q}  q |0 \rangle \, ,
\end{array}
\end{equation}
where $q$ refers to the up and down quark, $e_c$ is the quark charge. 
The $\chi$, $\kappa$ and $\xi$ in Eq. (5) are the quark condensate 
susceptibilities, which have been the subject of various 
studies \cite{IOFFE,Balit,Kogan,Chiu}. Their values employed by different 
groups are consistent with each other. We shall adopt the values
$\chi=-4.5\, \mbox{GeV}^{-2}$,  
$\kappa =0.4$, $\xi = -0.8$.  

\par
At the phenomenological level we have 
\begin{equation}\label{ph}
\mbox{Im} \Pi (s) = 
{1\over 4} \mu_{\Sigma_c} \lambda^2_{\Sigma_c} 
\delta^\prime (s-m_{\Sigma_c}^2) 
+ C \delta (s-m_{\Sigma_c}^2)  
+\mbox{Im} \Pi^{\mbox{pert}} (s) \theta (s-t_{\Sigma_c})
\end{equation}
where the first term corresponds to the $\Sigma_c$ magnetic moment and is 
of the double pole. The second term comes from the transition $\Sigma_c$ 
$\to$ excited states and is of single-pole. The third term is the usual 
continuum contribution and $t_{\Sigma_c}$ is the continuum threshold. 
The single-pole transition term does not damp out after Borel transform. 
So it should be explicitly included in the QSSR analysis. 

\par
Within an operator product expansion we obtain to lowest order of $\alpha_s$ 
and for condensates up to dimension six at the quark level,
\begin{equation}\label{q1}
\mbox{Im} \Pi (s) =\mbox{Im} \Pi^{\mbox{pert}} (s)
+\mbox{Im} \Pi^{\mbox{np}} (s) \, ,
\end{equation}
\begin{equation}\label{q2}
\mbox{Im} \Pi^{\mbox{pert}} (s) = {e_u\over 8} s (1-{m_c^2\over s})^3 \, ,
\end{equation}

\par
Making Borel tranform of $\Pi (p)$ and transferring the continuum contribution 
to the left hand side, we obtain:
\begin{equation}\label{sum}
\begin{array}{ll}
& 
M_B^2 \int_{m_c^2}^{t_{\Sigma_c}} \mbox{Im} \Pi^{\mbox{pert}} (s) 
e^{-{s\over M_B^2 }} ds L^{-{4\over 9}}
+\{ 
-{1\over 12}\chi e_u a^2  M_B^2 
(1-e^{-{t_{\Sigma_c}-m_c^2\over M_B^2}}) L^{-{16\over 27}} 
\\
&
-{1\over 24} e_c a^2 
-{1\over 36}  e_u a^2 ( {m_c^2 \over M_B^2} +1) 
+{1\over 96} \chi m_0^2 e_u a^2 ( 2{m_c^2 \over M_B^2} +1) L^{-{10\over 9}} 
\\
&
-{1\over 72} \kappa e_u a^2 ( 2{m_c^2 \over M_B^2} -1) 
-{1\over 36} \xi e_u a^2 ( {m_c^2 \over M_B^2} +1) 
\}  e^{-{m_c^2\over M_B^2}} L^{4\over 9}
\\
&= {1\over 4} 
(2\pi )^4 \lambda_{\Sigma_c}^2 e^{-\frac{m_{\Sigma_c}^2}{M_B^2}}
 \mu_{\Sigma_c} (1+C M_B^2) \, ,
\end{array}
\end{equation}
where 
$a=-(2\pi )^2 \langle 0 | {\overline  q}  q |0 \rangle 
=0.55   \mbox{GeV}^3 $, 
$a m_0^2 =(2\pi )^2 g_s \langle 0 | {\overline  q}\sigma\cdot G  q |0 \rangle $, 
$m_0^2 =0.8\mbox{GeV}^2$, $q=u$, $d$,  
$L=\frac{\ln (10M_B)}{\ln (5)}$. 
$C$ is the unknown constant to be determined from the sum rule, which 
parametrizes the transition contribution. 
We have checked that in the chiral limit $m_c \to 0$, our result reproduces 
the sum rule for nucleon magnetic moment \cite{IOFFE}. 

\par
For the charm quark mass we use  
$m_c =1.47\pm 0.1$GeV.
We adopt the estimated continuum threshold and overlapping amplititude 
in the QSSR analysis \cite{vmb,bag1,dai,pc,korner,ts}. 
$t_{\Sigma_c}=10$GeV$^2$ 
and 
$(2\pi )^4 \Lambda^2_{\Sigma_c}=0.8\pm 0.1$GeV$^6$. 
For the $\Sigma_c$ mass we may either use the predictions in the QSSR 
analysis or the experimental value \cite{DATA},   
$m_{\Sigma_c}=2.455$GeV.

\par
We may further improve the numerical analysis by taking into account of the 
renormalization group evolutions of the sum rule (\ref{sum}) through 
the anomalous dimensions of the condensates and currents. 
The working interval of the Borel mass $M_B^2$ 
for the sum rule (\ref{sum}) 
is $1.7 \mbox{GeV}^2 \leq M_B^2 \leq 2.5 \mbox{GeV}^2$ 
where both the continuum contribution 
and power corrections are controllable \cite{vmb,bag1}. 
Moving the factor 
${1\over 4}(2 \pi )^4 \lambda^2_{\Sigma_c} 
e^{- {m_{\Sigma_c}^2\over M_B^2}}$ on the right hand side to the left and 
fitting the new sum rules with a straight line approximation 
we may extract the $\Sigma_c$ magnetic moment. 
We show the Borel mass dependence of the new sum rule 
and the fitting straight line in Fig. 1 for the 
continuum threshold  
$t_{\Sigma_c}=10$GeV$^2$. 

\par
It can be seen in Fig. 1 that the nondiagonal transition  
contribution is important though it is not 
dominant in the working interval of the Borel mass  
$1.7 \mbox{GeV}^2 \leq M_B^2 \leq 2.5 \mbox{GeV}^2$. 
The sum rule is insensitive to the 
susceptibilities $\kappa$ and $\xi$ due to their small values. 
Their contributions are less than $5\%$. 
When $\chi$ varies from $-4.5$GeV$^{-2}$ to $-3.5$GeV$^{-2}$ or 
to $-5.5$GeV$^{-2}$, the sum rules change within $10\%$. 

\par
Our final result is $\mu_{\Sigma^{++}_c}=(5.4 \pm 0.5) {e\over 2m_{\Sigma_c}}$,  
where ${e\over {2 m_B}}$ is   
a natural unit in QSSR analysis of the baryon magnetic moment. 
By replacing $e_u$ in (\ref{sum}) with ${e_u +e_d \over 2}$ or $e_d$ 
we arrive at the magnetic moments for the other $\Sigma_c$ multiplets,  
$\mu_{\Sigma^{+}_c}=(0.6 \pm 0.1) {e\over 2m_{\Sigma_c}}$  
and 
$\mu_{\Sigma^{0}_c}=(-4.2 \pm 0.4) {e\over 2m_{\Sigma_c}}$.  
In unit of nuclear magneton 
$\mu_{\Sigma^{++}_c}=(2.1 \pm 0.3) \mu_N$, $\mu_{\Sigma^{+}_c}= (0.23 
\pm 0.03) \mu_N$ and   
$\mu_{\Sigma^0_c}=(-1.6\pm 0.2) \mu_N$.  

\par
Similarly we can extract the magnetic moments of $\Lambda_c$ with 
the following interpolating current. 
\begin{equation}
\eta_{\Lambda_c}(x) =\epsilon^{abc} [{u^a}^T (x) C\gamma_5 u^b(x) ] c^c(x) 
\end{equation}
The final sum rule reads as follows:
\begin{equation}\label{sum1}
\begin{array}{ll}
& 
{3e_c \over 16} M_B^2
\int_{m_c^2}^{t_{\Lambda_c}}  
[{m_c^2 \over 2} (1-{m_c^4\over s^2}) -{s\over 3}(1-{m_c^6\over s^3}) ]
e^{-{s\over M_B^2 }} ds L^{-{4\over 9}}\\
&
+\{ 
-{e_c \over 24}  a^2 
+{e_u +e_d \over 144}  a^2 ( 1- {m_c^2 \over M_B^2} ) \\
&
+{e_u +e_d \over 576} \chi m_0^2 a^2  L^{-{10\over 9}} 
+{e_u +e_d \over 48} \kappa  a^2  
\}  e^{-{m_c^2\over M_B^2}} L^{4\over 9}
\\
&= {1\over 4} 
(2\pi )^4 \lambda_{\Lambda_c}^2 e^{-\frac{m_{\Lambda_c}^2}{M_B^2}}
 \mu_{\Lambda_c} (1+C M_B^2) \, ,
\end{array}
\end{equation}
With the parameters \cite{vmb,bag1,dai,pc,korner,ts} 
$t_{\Lambda_c}=10$GeV$^2$, 
$(2\pi )^4 \lambda^2_{\Lambda_c}=1.0\pm 0.2$GeV$^6$   
and 
$m_{\Lambda_c}=2.285$GeV, 
we get 
$\mu_{\Lambda_c}=(0.15\pm 0.05) \mu_N$. The Borel dependence of 
$\mu_{\Lambda_c}$ and the fitting line is shown in Fig. 2. As in the  
analysis of the $\Sigma_c$ magnetic moment, the straight line approximation 
is good. 

\par
It is not difficult to extend the same analysis to extract the magnetic 
moments of $\Sigma_b$ and $\Lambda_b$. Yet the overlapping amplititudes 
$\lambda_{\Sigma_b}$ and $\lambda_{\Lambda_b}$ determined in the QSSR 
approach with finite bottom quark mass have large errors. So we do not 
tend to present numerical results here. 

\par
In summary, we have calculated the magnetic moment of $\Sigma_c$ and 
$\Lambda_c$ using the 
external field method in the QCD sum rules. 
There are no experimental data for the heavy baryon magnetic 
moments. Yet naive predictions have been made in the phenomenological models 
such as nonrelativistic quark model (NRQM) \cite{nr1,nr2}, bag model \cite{bm} 
and the Skyrme model \cite{sk1,sk2}. Especially in the quark model  
the heavy baryon magnetic moments have a rather simple form. 
$\mu_{\Lambda_c}=\mu_c$, 
$\mu_{\Sigma_c^{++}} ={8\over 9} \mu_p -{1\over 3}\mu_c$, 
$\mu_{\Sigma_c^{+}} ={2\over 9} \mu_p -{1\over 3}\mu_c$ 
and
$\mu_{\Sigma_c^{0}} =-{4\over 9} \mu_p -{1\over 3}\mu_c$.  
Our result of the $\Sigma_c$ magnetic moment is in good agreement with the 
NRQM prediction. In the $\mu_{\Lambda_c}$ sum rule (\ref{sum1}), the 
contribution from higher-dimension condensates is significant and comparable 
with the charm quark perturbative contribution numerically, since the 
light quark perturbative contribution and the induced light quark 
condensate vanishes. If we turn off all the higher-dimension nonperturbative 
contribution by setting the quark condensate $a=0$, we arrive at 
$\mu_{\Lambda_c}=0.35\mu_N$, which is very close to the NRQM prediction!
So the higher-dimension condensates lead to the possible deviation from 
the naive quark model result. It will be very interesting to measure 
$\mu_{\Lambda_c}$ experimentally.

\par
This work is supported in part by the National Natural Science Foundation
of China and the Doctoral Program of State Education Comimssion of China.  
It is also supported in part by the National Science Council of 
R.O.C. (Taiwan) under the grant NSC84-2112-M002-021Y.

\newpage
\vskip 1cm
\begin{center}
{\bf Figure Captions}
\end{center}

Fig. 1 The Borel mass dependence of the $\Sigma^{++}_c$ magnetic moment 
for the continuum threshold  
$t_{\Sigma_c}=10$GeV$^2$. 
The solid curve is the QCD sum rule prediction for $\mu_{\Sigma^{++}_c}$.  
The dotted line is a straight-line approximation. The intersect with 
Y-axis is the $\Sigma_c$ magnetic moment in unit of ${e\over 2m_{\Sigma_c}}$. 

\vskip 1cm

Fig. 2 The Borel mass dependence of $\mu_{\Lambda_c}$ and the fitting 
straight line. 

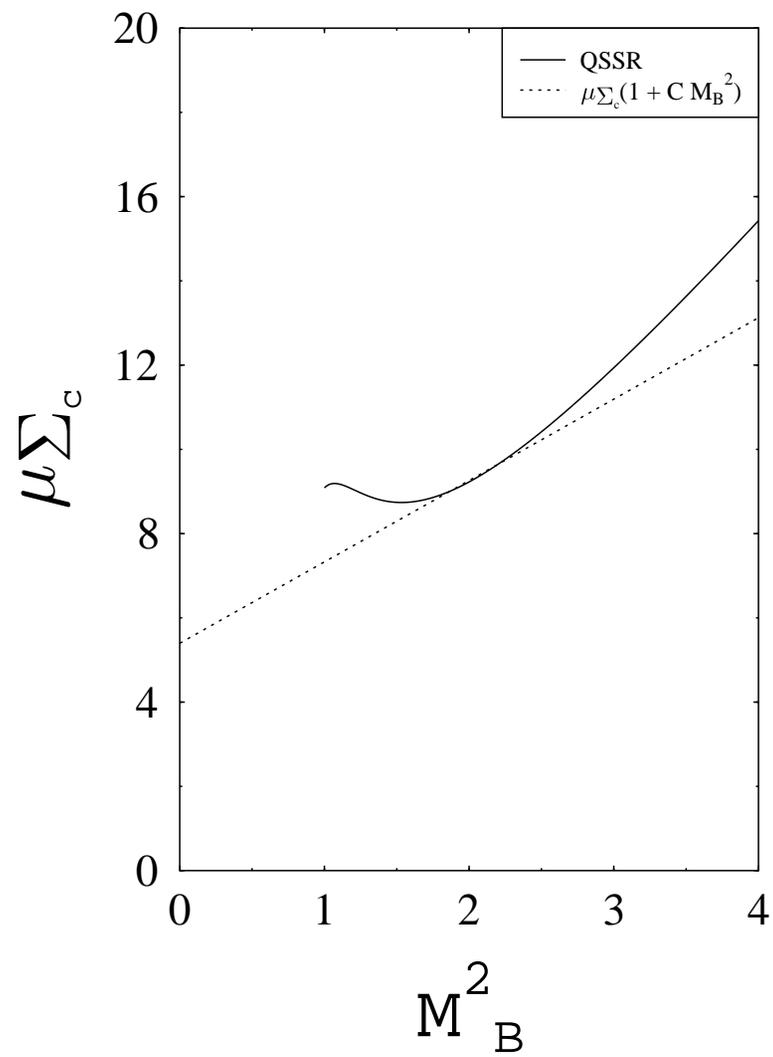
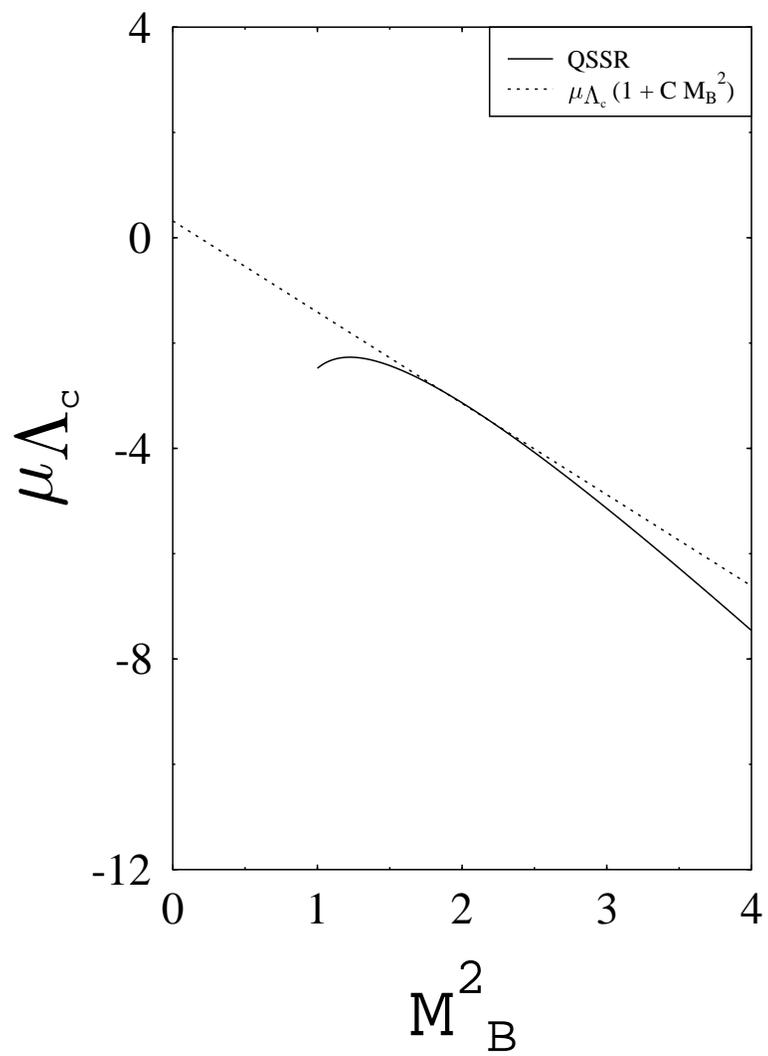
\end{document}